\documentclass[twocolumn,showpacs,prb,amsfonts,amsmath,amssymb,floatfix,groupedaddress]{revtex4}
\usepackage{color}
\usepackage{hhline}
\usepackage{mathrsfs}
\usepackage{graphicx}
\usepackage{dcolumn}
\usepackage{bm}% bold math
\usepackage{multirow}
\usepackage{booktabs}
\usepackage{afterpage}
\usepackage{amsmath}
\usepackage{ulem}

\begin{document}

\title{Stochastic formalism for thermally driven distribution frontier: A nonempirical approach to the potential escape problem}

\author{Ryosuke Akashi$^{1}$}
\thanks{akashi@cms.phys.s.u-tokyo.ac.jp} 
\author{Yuri S. Nagornov$^{1}$}

\affiliation{$^1$Department of Physics, The University of Tokyo, Hongo, Bunkyo-ku, Tokyo 113-0033, Japan}

\begin{abstract}
We develop a non-empirical scheme to search for the minimum-energy escape paths from the minima of the potential surface to unknown saddle points nearby. A stochastic algorithm is constructed to move the walkers up the surface through the potential valleys. This method employs only the local gradient and diagonal part of the Hessian matrix of the potential. An application to a two-dimensional model potential is presented to demonstrate the successful finding of the paths to the saddle points. The present scheme could serve as a starting point toward first-principles simulation of rare events across the potential basins free from empirical collective variables.
\end{abstract}

\maketitle

{\it -Introduction.}
Theoretical description of rare events concerns various issues in a wide range of research fields. A class of rare events involves transitions across the basins of the potential surface in the atomic configuration space, driven by thermal fluctuation; for example, folding of proteins, molecular reactions, diffusion of impurities in solids, nucleation, etc. There are roughly two types of interest in this context; how to efficiently sample distinct ``relevant" (e.g., in terms of the Boltzmann weight) states separated by the potential barrier, and how to identify the trajectory--the path that connect the states of interest. The present work focuses on the latter.

Among the possible paths connecting the states, the ones drawn from the intermediate saddle points (transition states) by tracking the steepest descent directions, often called reaction coordinates or minimum energy paths (MEPs), are of particular interest since they presumably dominate the target transition process~\cite{Fukui-ReacCoord1970}. Once the position of the saddle points or the neighboring final states are known, it is not so difficult to specify the trajectories along MEPs as well as to estimate their net probability of occurrence, as exemplified by the nudged elastic band~\cite{Henkelman-NEB-1, Henkelman-NEB-2} and transition path sampling~\cite{Dellago-TPS} methods. However, finding these states in the vast configuration space is formidable. Ascending the MEPs starting from a potential minimum is possible by utilizing the local Hessian matrix of the potential~\cite{Hilderbrandt-NewtonRaphson, Cerjan-Miller-eigenvec-follow} but practically difficult because it requires sensitive tuning of parameters~\cite{Cerjan-Miller-eigenvec-follow, Tsai-Jordan-eigenvec-follow-compare}.

\begin{figure}[t]
 \begin{center}
  \includegraphics[scale=0.7]{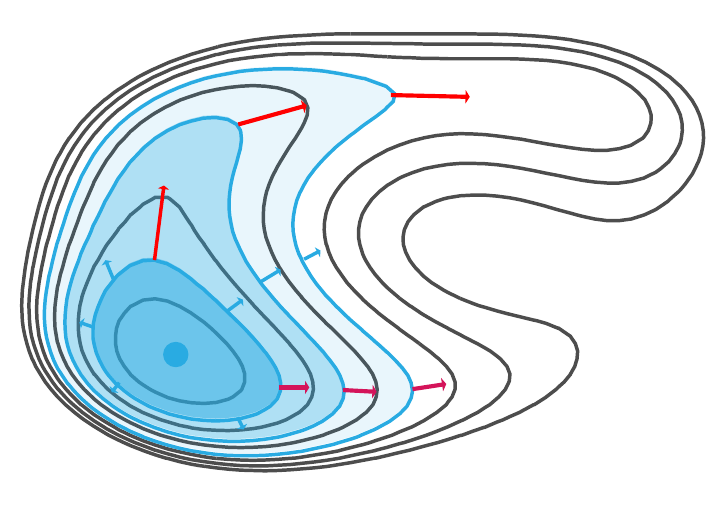}
  \caption{Temporal spread of the distribution function $p(x, t)$ under thermal fluctuation depicted by shades. The contour of the potential energy is depicted by solid lines. From the movement of the far-reaching frontiers we can in principle derive the reaction paths as indicated by red arrows.}
  \label{fig:escape-pic}
 \end{center}
\end{figure}

In the current standard atomistic simulation methods such as molecular mechanics and molecular dynamics, the escaping trajectories and free-energy landscape along them are calculated by applying artificial force and/or using {\it a priori} knowledge of collective variables that well characterize the target processes. Numerous celebrated methods are of this type; although we cannot append the exhaustive list, let us exemplify a few. A class of methods utilizes additional potential force; umbrella sampling~\cite{Torrie-Valleau-umbrella-JCP1977, Frenkel-umbrella-free-ene}, steered dynamics,~\cite{SteeredMD-orig-Sci1996} hyperdynamics~\cite{Voter-hyperdyn-JCP1997,Voter-hyperdyn-PRL1997}, metadynamics~\cite{Laio01102002}, adaptive biasing force~\cite{Darve-ABF-JCP2001}, hyperspherical search~\cite{Ohno-Maeda-ADDF2004}, and artificial force induced reaction~\cite{Maeda-Morokuma-AFIR2010} methods. Other methods concern constraining the microscopic dynamics within isosurfaces of the collective variables; the blue-moon~\cite{Carter-Ciccotti-bluemoon,Sprik-Ciccotti-bluemoon} and targeted dynamics~\cite{Schlitter-Wollmer-targetedMD} methods. An alternative that does not involve the artificial force is to run multiple simulations with coupling to the thermostat and select some of them according to a sampling measure defined with the collective variables; forward-flux sampling~\cite{Allen-tenWolde-FFS-PRL2005}, parallel cascade~\cite{Harada-Kitao-PaCSMD-JCP2013} and structural dissimilarity sampling~\cite{Harada-Shigeta-SDS-JCTC2017,Harada-Shigeta-SDS-JCC2017}. With these strategies, however, one always suffers from possible warping of the trajectories due to the additional force and/or insufficient sampling due to bad correspondence between the empirical collective variables and the MEPs~\cite{note-motivation, note-PRD}. A fundamental method to track the MEPs free from the above-mentioned problems is desired, especially for the simulation of systems whose experimental reference is not available.

In this letter, we propose an efficient method to generate the trajectory that ascends the MEPs without resorting to the collective variables. The point is that the MEPs guide the maximally probable trajectories from the target initial states under the thermal fluctuation. A natural way to specify them is therefore to refer to the time-dependent conditional probability, as it is directly related to the probable escape events (Fig.~\ref{fig:escape-pic}); $p({\bm x}, t| {\bm x}_{\rm 0}, 0)$, with ${\bm x}_{\rm 0}$ set near a potential minimum. We develop a stochastic walker-type algorithm that generates the dynamics of the spreading frontier of $p({\bm x}, t| {\bm x}_{\rm 0}, 0)$ to the saddle points, with which the escaping trajectories through the correct MEPs are realized. This algorithm only uses the local values of the potential gradient and {\it diagonal part} of the Hessian matrix, does not introduce prior definitions of the collective variables, and can therefore enable us the non-empirical search for the MEPs and final states beyond them. %Our method is, also, in principle applicable to quantitative evaluation of the time-dependent escape probability, though discussions on this aspect is left for later studies. %Since the rarity of such transitions is presumably dominated by the height of the potential barrier, the paths between the target configurations where the barrier height has the lowest values have been of special interest as reaction coordinates (RCs).

%The problem of this class of rare transitions is composed of two issues; specifying the RC(s) and evaluating the probability of occurrence. For the former, the task is rendered simpler when the final and/or transition states are known. 

%By limiting the scope to the former issue, it is generally done either by applying additional potential to render the potential slope upward to downward to the transition state. With this strategy, however, the total potential surface is warped and the trajectories in the simulations do not strictly agree with the ideal reaction paths. Another way to generate the trajectories along the RCs is to select the configurations under random force representing the thermal effect according to some sampling measures. This type of methods need prior assumption on the ``plausible" RCs along which the measure is defined. The prior assumption on the RCs also concerns the former methods.

%With these methods,  the trajectories realized in the simulation is unavoidably warped from the ideal one. The methods with additional force

% need prior assumption on the ``plausible" reaction coordinate along which the potential force is exerted or the sampling measure is defined. With this assumption
%The former Specifying the RCs in the configuration space is, however, quite a nontrivial problem. If the final and/or transition states are known, the problem is simpler and some established methods

{\it -Basic consideration.} 
We start with a general discussion on the system under effects of the potential force and coupling to the thermal bath as a random force, described by the following Langevin equation (in Ito's convention)~\cite{Gardiner-book}
\begin{eqnarray}
dx_{i} &=& \frac{p_{i}}{m}dt; \\
dp_{i} &=& \left[-\partial_{i} U({\bm x}) -\frac{\Gamma }{m}p_{i} \right]dt + \sqrt{2\Gamma k_{\rm B}Tdt}W_{i}
,
\end{eqnarray}
where $U({\bm x})$, $m$, $k_{\rm B}$ and $T$ are the potential, mass of the particles, Boltzmann constant, and temperature, respectively. $i$ is the index for the degrees of freedom and $\partial_{i}\equiv\frac{\partial}{\partial x_{i}}$. ${\bm W}$ is the vector whose components are randomly generated from the standard normal distribution at each step. $\Gamma$ is the friction constant. Note that this is the very formula employed in the Langevin molecular dynamics simulation. We eliminate the fast variable ${\bm p}$ for simplicity and get to~\cite{Gardiner-book}
\begin{eqnarray}
dx_{i} =  -\frac{\partial_{i} U({\bm x})}{\Gamma} dt+ \sqrt{\frac{2k_{\rm B} Tdt}{\Gamma}}W_{i}
.
\label{eq:SDE-x}
\end{eqnarray}
This form directly relates to the molecular mechanics.
%For simplicity we here work with this level; but note that the later discussions can obviously be applied with ${\bm p}$ retained, apart from practical complications.

There are two ways of describing the stochastic dynamics; one is the equation with random terms for an individual particle (walker) whose state is characterized by ${\bm x}$, and the other is the deterministic equation for the distribution of the walkers $p({\bm x}, t)$. The latter counterpart of Eq.~(\ref{eq:SDE-x}) is the Smoluchowski equation~\cite{Gardiner-book}
\begin{eqnarray}
\partial_t p({\bm x},t) &=& \frac{1}{\Gamma}\partial_i [ (\partial_{i}U({\bm x})) + k_{B}T \partial_i]p({\bm x},t)
\\
&\equiv& \hat{L}_{\rm Sm}p({\bm x},t)
.
\label{eq:Smoluchowski}
\end{eqnarray}
Hereafter the product with the identical index ($\dots A_{i} B_{i} \dots$) implies summation with respect to that. Note that this equation has the Boltzmann distribution $p_{\rm eq}({\bm x})\propto {\rm exp}[-U({\bm x})/k_{\rm B}T]$ as the stationary solution. 

\begin{figure}[h]
 \begin{center}
  \includegraphics[scale=0.32]{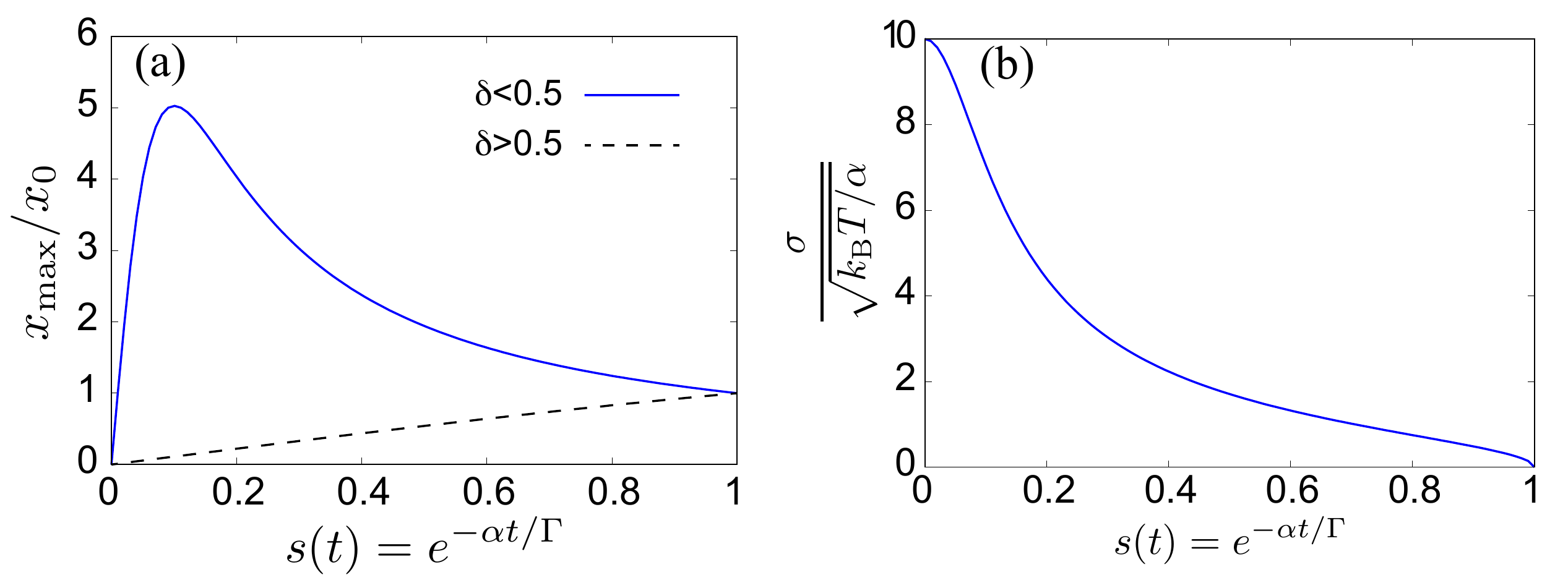}
  \caption{Behavior of function $q(x,t)$ [Eq.~(\ref{eq:OU-q})]. (a) Time evolution of the center of $q(x,t)$ with $\delta$=0.01 (solid line) and 0.9 (dashed line), respectively. (b) Time evolution of the width of $q(x,t)$ with $\delta$=0.01.}
  \label{fig:OUprocess}
 \end{center}
\end{figure}

Let us next consider how the distribution evolves toward $p_{\rm eq}({\bm x})$. Starting from the initial distribution $p({\bm x}, 0)=\delta({\bm x}-{\bm x}_{0})$ with ${\bm x}_{0}$ near the potential minimum, $p({\bm x}, t)$ gradually spreads away. The extent of the spread then reflects the relative height of the potential; $p({\bm x}, t)$ tends to spread far to the direction of potential ``valleys"--where the slope of the potential is small (Fig.~\ref{fig:escape-pic}). We can roughly imagine that the structure of the valley paths from the initial point can be derived by tracking the dynamics of the ``frontier" of the spread.

To substantiate this idea we first analyze the Orenstein-Uhlenbeck process in one dimension
\begin{eqnarray}
\partial_t p(x,t) = \frac{1}{\Gamma}\partial_x (\alpha x + k_{B}T \partial_x)p(x,t)
.
\end{eqnarray}
This is the equation of the distribution of the walkers under parabolic potential $U(x)=\frac{1}{2}\alpha x^2$ and the thermal fluctuation. The analytic solution of $p(x, t)$ is given as
\begin{eqnarray}
p(x,t)
=
\sqrt{\frac{\alpha}{2\pi k_{\rm B}T (1-s^2)}}
{\rm exp}
\left[
-
\frac{\alpha(x-sx_{0})^2}{2k_{\rm B}T(1-s^2)}
\right]
\end{eqnarray}
with $s\equiv s(t)={\rm exp}(-\alpha t/\Gamma)$. According to this form, in a short time where $\alpha t/\Gamma$ is small, the drift of the center of $p(x,t)$ from $x_{0}$ is $O(t)$ whereas the spread of $p(x,t)$ is $O(\sqrt{t})$, implying that the short-time behavior is more like the Wiener process, the Brownian motion under zero potential force. This fact derives an intriguing property of $p(x,t)$. If we factorize $p(x,t)$ with the equilibrium distribution, which definitely reflects the potential height, as $p(x,t)\propto p_{\rm eq}(x) q(x,t)$, the remaining factor $q(x,t)$ which reflect the short-time Wiener-like property, should have larger value for $x$ with larger $U(x)$. This reasoning yields that the conditional probability distribution with the starting point $x_{0}$ on a frontier of the original $p(x,t)$ in the middle of the potential slope will, if factorized by $p_{\rm eq}(x)$, drift farther up the slope.

The above expectation is verified by the analytic form of $q(x,t)$.  Defining it by $p(x,t)={\rm exp}[-\alpha (1-\delta)x^{2}/2]q(x,t)$ with parameter $\delta$ for generalization, we get
\begin{eqnarray}
{\rm ln}q(x,t)
=
f(t)-\frac{1}{2\sigma^2(\delta; s)}(x-x_{\rm max}(\delta; s))^2
\label{eq:OU-q}
\end{eqnarray}
with $f(t)$ being the terms depending only on $t$, and
\begin{eqnarray}
\sigma^2(\delta; s)
=\frac{k_{\rm B}T}{\alpha} \frac{1-s^2}{\delta+(1-\delta)s^2},
\\
x_{\rm max}(\delta; s)
=
x_{0} \frac{s}{\delta + (1+\delta)s^2}
.
\end{eqnarray}
Note that $q(x,t)$ keeps the Gaussian form at any $t$ like $p(x,t)$. Time evolution of its maximum position $x_{\rm max}$ and spread $\sigma$ is depicted in Fig.~\ref{fig:OUprocess}. When $\delta < 1/2$, $x_{\rm max}$ first goes upward the potential surface, reaches to the maximum $x_{0}/\sqrt{\delta(1-\delta)}$, and finally comes back to the potential minimum. On the other hand, $\sigma$ continuously increases toward $\sqrt{k_{\rm B}T/\delta\alpha}$. This result indicates a dual character of $q(x,t)$ for $\delta <1/2$ depending on the initial position $x_{0}$. When $x_{0}$ is set well apart from the minimum, the whole distribution continues to go upward; when it is near to the minimum, on the other hand, it just gradually spreads around the minimum as the change of $x_{\rm max}$ is invisibly smaller than $\sigma$. The threshold length scale is $x_{0}\sim \sqrt{k_{\rm B}T/\alpha}$.

This character of $q(x,t)$ is apparently utilizable for finding the paths through the valleys to the saddle points of the potential surface in higher dimensions. Suppose the initial position ${\bm x}_{0}$ is set at the middle of any of the valley paths. The distribution center will then go farther to the direction along that valley path, whereas it will keep its position invariant in the other directions in which the potential is presumably parabolic.

{\it -Stochastic walker algorithm.}
In view of the application to the escape problem from the potential basins, we then construct a microscopic stochastic algorithm to reproduce $q({\bm x},t)$. From the Smoluchowski equation [Eq.~(\ref{eq:Smoluchowski})], the corresponding equation for $q({\bm x},t)$ with general transformation $p({\bm x},t)={\rm exp}[-V({\bm x})/k_{\rm B}T]q({\bm x},t)$ is given by~\cite{Giardina-review-JStatPhys2011} $\partial_{t}q({\bm x},t)=\hat{L}'q({\bm x},t)$ with
\begin{eqnarray}
\hat{L}'
&=&
e^{V({\bm x})/(k_{\rm B}T)}\hat{L}_{\rm Sm} e^{-V({\bm x})/(k_{\rm B}T)}
\nonumber \\
&=&
\frac{1}{\Gamma}
\partial_{i} [\partial_{i} (U({\bm x})-2V({\bm x}))]+\frac{k_{\rm B}T}{\Gamma} \partial_{i}^{2} + \frac{F({\bm x})}{\Gamma}
,
\\
F({\bm x})
&=&
\partial_{i}^{2}V({\bm x})
+
\frac{1}{k_{\rm B}T}(\partial_{i}V({\bm x}))[\partial_{i}(V({\bm x})-U({\bm x}))]
.
\label{eq:Ffunc}
\end{eqnarray}
We here reformulate this so that the conservation of $q({\bm x},t)$ ($\int dx q({\bm x},t)=const.$) is assured; by redefining $q({\bm x},t)$ with a time-dependent coefficient $C(t)$ by
\begin{eqnarray}
p({\bm x},t)
=C(t)e^{-V({\bm x})/(k_{\rm B}T)}q({\bm x},t)
,
\label{eq:q-factorize}
\end{eqnarray}
we get
\begin{eqnarray}
\partial_{t} q(x,t)
\equiv
[\hat{L}'_{\rm Sm}+\hat{L}_{\rm rate}] q(x,t)
\end{eqnarray}
with
\begin{eqnarray}
\hat{L}'_{\rm Sm}
=
\frac{1}{\Gamma}
\partial_{i} [\partial_i (U({\bm x})-2V({\bm x}))]+\frac{k_{\rm B}T}{\Gamma}\partial_{i}^2,
\\
\hat{L}_{\rm rate}
=
\frac{1}{\Gamma}
\left[
F({\bm x})-\langle F\rangle_{q({\bm x},t)}
\right],
\\
\partial_t {\rm ln}C(t) = \frac{1}{\Gamma}\langle F\rangle_{q({\bm x},t)}
.
\end{eqnarray}
Here we define the average of the function $f({\bm x})$ by $\langle f\rangle_{q({\bm x},t)} = \int d{\bm x} q({\bm x},t)f({\bm x})$. %This treatment conserve the number of walkers on average and does in spirit the same thing as the recently proposed strictly number-conserving cloning algorithm.~\cite{Brewer-Jack-Nconserve-Clone}

The stochastic time-evolution process for individual walkers, whose assembly reproduces $q({\bm x},t)$, is then formulated. The evolution by timestep $\tau$ is formally represented as
\begin{eqnarray}
q(x,t+\tau)
={\rm exp}\{[\hat{L}'_{\rm Sm}+\hat{L}_{\rm rate}] \tau\}q(x,t)
.
\label{eq:q-evolve}
\end{eqnarray}
The operation ${\rm exp}\{\hat{L}'_{\rm Sm} \tau\}$ on the distribution function is recast to the Langevin equation~\cite{Gardiner-book} [Eq.~(\ref{eq:SDE-x})] with potential modified to $U-2V$ for the walkers. To utilize this, we apply the Suzuki-Trotter decomposition~\cite{Suzuki-Trotter-1-AMS1959,Suzuki-Trotter-2-CMP1976} ${\rm exp}\{[\hat{L}'_{\rm Sm}+\hat{L}_{\rm rate}]\tau \}\simeq {\rm exp}[\frac{\hat{L}_{\rm rate}\tau}{2}]{\rm exp}[\hat{L}'_{\rm Sm}\tau] {\rm exp}[\frac{\hat{L}_{\rm rate}\tau}{2}]+O(\tau^3)$. The whole time evolution operation is then implemented as successive steps of simple multiplication of the factor ${\rm exp}[\frac{\hat{L}_{\rm rate}\tau}{2}]$ (recast to replicating/removing the walkers ${\bm x}$ by the corresponding probability; importance sampling~\cite{Grassberger-importance-sampling-PRE1997}) and the Langevin evolution of the walkers. With the factorization of $C(t)$ in Eq.~(\ref{eq:q-factorize}) the total number of walkers $N_{\rm w}$ is conserved on average. Our foundation has been inspired by the construction of the diffusion Monte Carlo method~\cite{RevModPhys.73.33}.

Although $V({\bm x})$ can be set arbitrarily, as a useful form, we propose to set $V({\bm x})=(1-\delta)U({\bm x})$. This form definitely reflects the convex/concave structure of $U({\bm x})$ and therefore we can expect the trajectories ascending the MEPs of $U({\bm x})$. This setting is convenient because, in most situations of interest, the value of $U({\bm x})$ for a given ${\bm x}$ is available through a formula or microscopic calculations. Another advantage is that the algorithm is executable with only the diagonal part of the Hessian matrix of $U({\bm x})$ [Eq.~(\ref{eq:Ffunc})], in contrast to the preceding deterministic methods that require the whole matrix~\cite{Hilderbrandt-NewtonRaphson, Cerjan-Miller-eigenvec-follow}. 

Here we summarize our algorithm to search for the MEPs. (I) Generate initial ``frontier" distribution of the walkers $q({\bm x},0)\simeq \delta({\bm x}-{\bm x}_{0})$ by executing usual Langevin dynamics with potential $U({\bm x})$ for some duration at a temperature $T_{\rm ini}$ and selecting some walkers reaching far from the known minimum of $U({\bm x})$. Afterwards, (II) Setting the temperature $T_{\rm esc}$ ($<T_{\rm ini}$), execute the time evolution (Eq.~(\ref{eq:q-evolve})) with $V({\bm x})=(1-\delta)U({\bm x})$. Representative points (e.g. maximum) of the resulting distribution of $q({\bm x}, t)$ or $p({\bm x},t)$ draw the trajectories that go to the neighboring saddle points. The factor $\sqrt{T_{\rm esc}/\delta}$ dominates the ideal spread of $q({\bm x},t)$ and therefore controls the number of walkers $N_{\rm w}$ required for stable calculations: setting this factor small, the whole shape of $q({\bm x},t)$ can be represented with small $N_{\rm w}$, but its behavior could be subject to outlier walkers departing normal to the MEPs, as shown later. Note that the stable simulation is even then achieved in the small $\tau$ limit.

Although the resulting $q({\bm x},t)$ and $p({\bm x},t)$ have well-defined meaning as time-dependent conditional probability of ${\bm x}$ coupled to the thermal bath of $T_{\rm esc}$ with strength $1/\Gamma$, in this work, we just exploit them to derive the MEPs and do not address its quantitative aspect as absolute escape probability in nonequilibrium processes. We here simply regard $T_{\rm esc}$, $\Gamma$ and $\delta$ as fine-tuning parameters to stabilize the behavior of the simulation. %The latter aspect, which should have close relation to the recent study on large deviation function~\cite{}, will be left for future studies.

\begin{figure}[t]
 \begin{center}
  \includegraphics[scale=0.15]{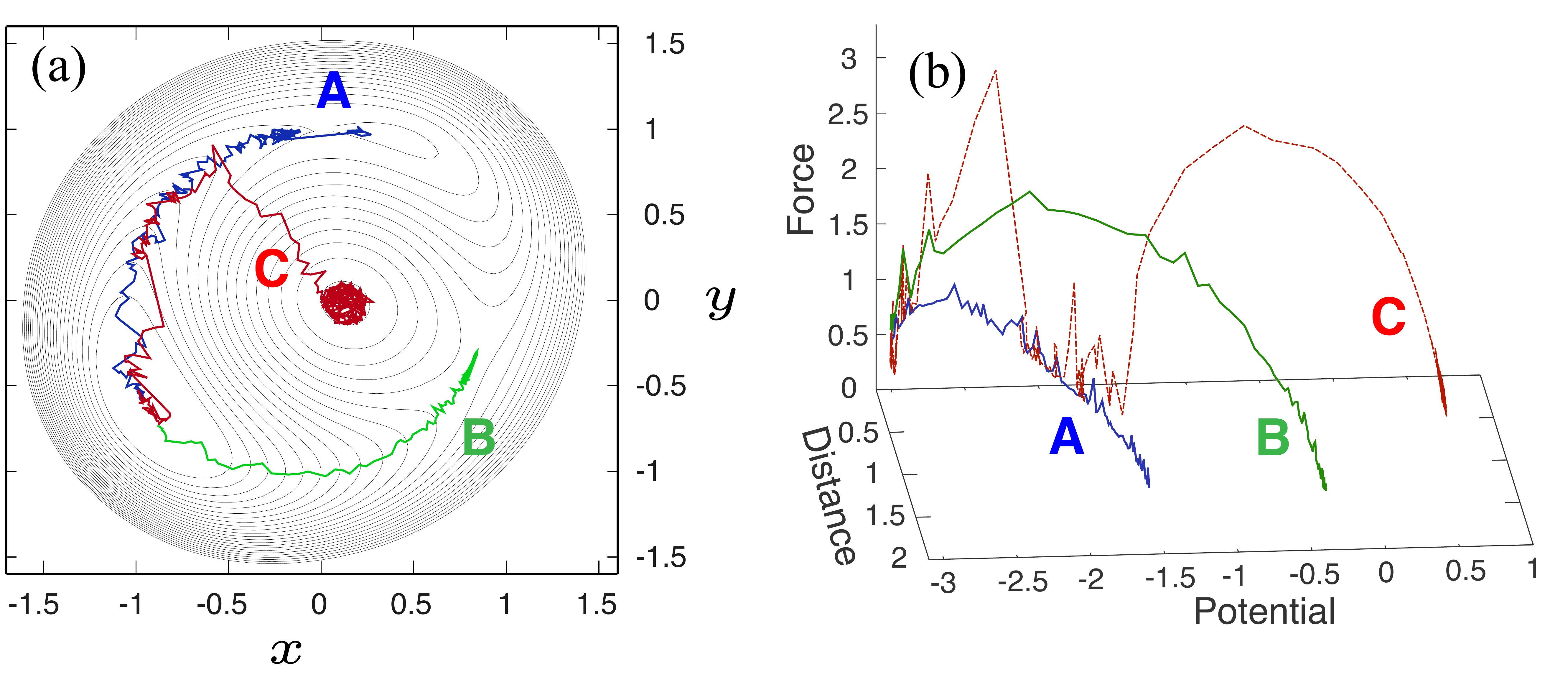}
  \caption{Finding the minimum energy path for a two dimensional model potential $U(x,y)$. (a) Contour plot of $U(x,y)$ with three representative trajectories: to the first and second saddle points and to the maximum. (b) Representation of the trajectories in terms of distance from the minimum, absolute value of the force, and the potential height, where the stationary (minimum, saddle, and maximum) points can be identified by the weak force.}
  \label{fig:path}
 \end{center}
\end{figure}

{\it -Application to a two-dimensional model.}
We show an application of the present algorithm to a two-dimensional model potential:
$U(x,y)
=2(x^2+y^2-1)(x^2+y^2)+\frac{1}{2}{\rm exp}(-x^2 y^2)+x-xy,$
which has a maximum near $(x, y)\simeq(0;0)$ and two minima: global minimum near (-0.9;-0.6) and local one near (0.4;0.9). 
This is a simple construction of the potential surface having non-linear MEPs (first and second terms) with subtle modification (third and fourth terms) as shown in Fig.~\ref{fig:path}(a). The number of walkers $N_{\rm w}$ and timestep $\tau$ were set to 200 and $5\times 10^{-4}$, respectively, where $N_{\rm w}$ somehow deviated from the original value due to the importance sampling steps. The temperatures $T_{\rm ini}$ and $T_{\rm esc}$ were 10$^{-2}$ and $8\times 10^{-3}$, whereas the $\delta$ parameter that defines the biasing potential $V=(1-\delta)U$ was $2\times 10^{-3}$. The friction constant $\Gamma$ was set to 10, whereas the duration time of the modified dynamics was $5\times 10$. 

Starting from the minimum $(x, y)\simeq(-0.9; -0.6)$, we executed 100 trials of the above-mentioned procedure and recorded the peak position of $p(x,y,t)$.~\cite{note-p-traj} For typical behavior of $q(x, y, t)$ and $p(x,y,t)$, see Supplemental videos. The representative resulting trajectories are shown in Fig.~\ref{fig:path}(a). We obtained the trajectories going through the first and second minimum energy paths to the two saddle points (A and B, respectively), depending on the initial guess of ${\bm x}_{0}$. The path search sometimes failed; during the simulation the walkers stray normal to the MEP and move up the slope to the maximum, as represented by trajectory C. This failure is understandable from Fig.~\ref{fig:OUprocess} as the case where a large fraction of walkers are accidentally driven normal to the valley. Seven failures of 100 trials were observed with $T_{\rm esc}=8\times 10^{-3}$ and $\delta=2\times 10^{-3}$, but we have confirmed that the failure rate can be reduced by tuning the parameters. The distance-force-potential plot in Fig.~\ref{fig:path}(b) can help to discriminate the different paths to transition states, as well as to detect the artificial departure like C; the latter appears with drastic change in the direction. This plot obviously applies to higher dimensions.

{\it -Discussion and future perspectives.}
In this letter, we have proposed a non-empirical scheme to generate the minimum-energy escape paths by tracking the time evolution of the biased conditional probability under weak thermal fluctuation. Numerical methods to treat the dynamics with probability bias have been widely applied recently for evaluating the large deviation function of physical quantities~\cite{PhysRevLett.96.120603, PhysRevLett.112.090602, PhysRevE.93.062123} and generating rare trajectories~\cite{Tailleur-Kurchan-Lyapnov-weighted}, though these applications mainly concern the system where the quantities to be biased are given or targeted {\it a priori}. Our demonstration shows that the biasing by the factor using the potential function $U({\bm x})$ itself can be utilized to extract the spatial coordinates that well represent the dominant escape trajectories from high-dimensional configuration space.

A notable thing is that the biased $q({\bm x}, t)$ is of localized form with width $O(\sqrt{T_{\rm esc}/\delta})$. Our expectation is that the curse of dimensionality, which causes the exploding $N_{\rm w}$ required for reliable calculation in usual walker-type methods, could be mitigated thanks to the localization of $q({\bm x}, t)$. Applications to larger and realistic systems are under way. The narrow extent of $q({\bm x}, t)$ also suggests that it accurately reproduces the original $p({\bm x}, t)$ via Eq.~(\ref{eq:q-factorize}) at least for the region where the walkers are distributed. The present methodology could therefore provide a basis for estimating the absolute escape probability in non-equilibrium situations, which will be addressed in later studies.

Extensions for combination to the molecular dynamics simulation is an intriguing issue. Formally it is done by keeping the fast variable ${\bm p}$. Considering time-dependent $V({\bm x}, t)$ for the transformation in Eq.~(\ref{eq:q-factorize}), the present scheme could also be combined to the methods with time-dependent adaptive potentials such as metadynamics~\cite{Laio01102002}.  

\begin{acknowledgments}

%\acknowledgment
This research was supported by MEXT as Exploratory Challenge on Post-K computer (Frontiers of Basic Science: Challenging the Limits). This research used computational resources of the K computer provided by the RIKEN Advanced Institute for Computational Science through the HPCI System Research project (Project ID:hp160257, hp170244).

\end{acknowledgments}

\bibliography{reference}

\begin{thebibliography}{36}
\bibitem[{not({\natexlab{a}})}]{note-motivation}
\bibinfo{note}{Note that such methods are sometimes employed only to sample the
  states efficiently for calculating thermodynamics quantities, not the
  trajectories; the incorrect trajectories are then not regarded as a problem.}
\end{thebibliography}

\end{document}